\documentclass[11pt]{article} 

\usepackage[utf8]{inputenc}
\usepackage{amsmath, amssymb}
\usepackage{graphicx}
\usepackage{cite}
\usepackage{hyperref}
\usepackage[top=1in, bottom=1in, left=0.9in, right=0.9in]{geometry} 
\usepackage{titlesec} 

\titleformat{\section}{\normalfont\normalsize\bfseries}{\thesection}{1em}{} 
\titleformat{\subsection}{\normalfont\normalsize\itshape}{\thesubsection}{1em}{} 
\titlespacing*{\section}{0pt}{1ex plus .5ex minus .2ex}{0.8ex}
\titlespacing*{\subsection}{0pt}{0.8ex plus .4ex minus .2ex}{0.6ex}

\title{High-energy electron-beam induced defect engineering of monolayer MoS$_2$ for tunable optical properties}
\author{
  Anagha Gopinath$^{1}$,
  Faiha Mujeeb$^{2}$, 
  Subhabrata Dhar$^{2}$,
  Jyoti Mohanty$^{1}$\thanks{Email: jmohanty@phy.iith.ac.in}   
}

\date{}
\begin{document}
\maketitle
\begin{center}
$^{1}$Nanomagnetism and Microscopy Laboratory, Department of Physics, Indian Institute of Technology Hyderabad, Kandi, Sangareddy, Telangana, India \\
  $^{2}$Department of Physics, Indian Institute of Technology Bombay, Powai, Mumbai, India\\
\end{center}
\begin{abstract}
Structural defects in 2D-transition metal dichalcogenides are critical in modulating their optical and electrical behavior. Nevertheless, precise defect control within the monolayer regime poses a significant challenge. Herein, a high-energy (1MeV) electron beam irradiation strategy is harnessed to induce defects in monolayer MoS$_2$. Controlled variation of electron-beam irradiation time tunes the defect density, as reflected by the evolution of defect-mediated photoluminescence characteristics. The optically active defect emission appearing at $\approx$200-300meV below the A exciton at 85K exhibits a systematic increase in intensity with prolonged exposure and saturates at higher laser excitation power. Circular polarization-resolved photoluminescence spectroscopy reveals strong suppression of valley polarization of the A exciton after irradiation. Complementary x-ray photoelectron spectroscopy identifies enhanced Mo-O bonding signatures in MoS$_2$ following irradiation. Kelvin probe force microscopy indicates the transition to p-type doping behaviour. A detailed temperature and power-dependent photoluminescence measurements further elucidate the optical behaviour of these defect states. Density functional theory calculations using these configurations establish that the transition between the conduction band and acceptor states within the bandgap accounts for the defect emission. This work presents a tunable route for defect engineering in monolayer TMDs, enabling controlled tailoring of their structural and optical properties for optoelectronic, electronic and valleytronic applications. 
\end{abstract}

\section{Introduction}
The growing demand for compact and high-performance technologies in quantum devices, optoelectronics and electronic device applications has positioned 2D materials as compelling candidates owing to their remarkably tunable properties~\cite{1,2,3,4,5,6}. Based on the tunable electronic and optical properties, 2D transition metal dichalcogenides (TMDs) have garnered considerable research interest in recent years~\cite{7,8,9}. They exhibit large absorption coefficients~\cite{10,11}, valley-selective optical selection rules~\cite{12,13}, and strong exciton binding energy arising out of strong light-matter interactions~\cite{14,15,16,17}. Along with these advantages, strong spin-orbit coupling~\cite{18}, high mobility, better electrostatic control ~\cite{20,21} and mechanical stability ~\cite{22}of TMDs make them promising for integrating into device architectures~\cite{23,24,25}. To further expand their functionalities, several strategies have been pursued, with defect engineering emerging as particularly effective. Structural defects like point defects, line defects and adatoms are inevitable in TMDs and significantly impact their electrical~\cite{26}, optical~\cite{27,28}, magnetic~\cite{29}, and electronic properties~\cite{58}. These defects can form intrinsically during the growth or be introduced extrinsically via external perturbations. In light of the experimental reports, a native defect density of $\sim$10$^{13}$cm$^{-2}$ was observed in monolayer TMDs that primarily arose from chalcogen vacancies~\cite{30,31}. The defect density has been modulated through diverse strategies such as plasma exposure~\cite{32}, chemical functionalization~\cite{33,34}, thermal annealing~\cite{35,36} and ion~\cite{37} and electron beam irradiation~\cite{38}. The controlled introduction of defects has yielded novel functionalities such as spin qubits~\cite{39}, single photon emitters~\cite{40}, improved thermoelectric efficiency~\cite{41}, and enhanced sensing behavior~\cite{42}. Despite these advantages, excess defect density reduces the chemical sensitivity and electrical conductivity and impairs device performance~\cite{43}. A comprehensive understanding of defect–property relationships is therefore essential for gaining fundamental insights and optimizing TMDs-based device performance.\\

Particle irradiation offers a precise ex-situ approach for defect engineering in 2D-TMDs and enables fine control over defect type and density by modulating the particle type, irradiation energy, and fluence~\cite{28}. The resulting irradiation-induced defects reshape the electronic, catalytic, optical and magnetic properties of TMDs~\cite{53,54,55,56,57}. In particular, atomic vacancies strongly influence the optical emission and absorption characteristics of TMDs through the formation of midgap states within the band gap~\cite{44}. Previous experiments have showcased photoluminescence (PL) quenching~\cite{45} and enhancement~\cite{46} after irradiation. Irradiating few-layer MoS$_2$ with protons increased the PL yield and suppressed the indirect transitions~\cite{47}. An enhancement in defect density and trion emission was reported in monolayer WS$_2$ exposed to gamma ray irradiation~\cite{48}. Room temperature defect-bound emission was observed in the PL spectra of WS$_2$ after controlled argon plasma treatment~\cite{49}. Furthermore, focused helium ion beam~\cite{50,51} and proton beam~\cite{52} generated optically active defect centres with sharp emission lines in h-BN encapsulated TMDs, which can serve as a host for single photon emitters. However, ion irradiation can lead to unintended lattice damage and potential ion implantation. Since electrons possess lower momentum, irradiating with electrons can be a better alternative for precisely creating defects in low-dimensional materials through knock-on damage, electronic excitations or ionization~\cite{38}. Despite its potential, previous works using electron beams have been limited to low-energy electron beams generated by transmission electron microscopy (TEM) and scanning electron microscopy (SEM). Studies have described the agglomeration of chalcogen vacancies into line defects under prolonged exposure to an 80 kV HRTEM beam~\cite{59,60}. Feng \textit{et al} experimentally demonstrated that the chalcogen vacancies induced in monolayer MoS$_2$ following 3kV electron beam exposure lead to enhanced photoresponse behavior and a reduced electron mobility~\cite{61}. Giubileo \textit{et al} reported a three-order increase in transistor channel current succeeding 10keV electron beam irradiation~\cite{62}. Even though a few reports have addressed the effect of high-energy ($\sim$MeV) electron beams on the electrical transport properties of monolayer TMDs~\cite{63,64}, systematic investigations of their influence on optical properties remain scarce. Recent work by Dash et al. reported sharp defect-bound emissions and single photon emitters in h-BN encapsulated MoS$_2$ under ultra-low energy electron beam irradiation~\cite{65}. However, the impact of high-energy ($\sim$MeV) electron beam induced defects on the structural stability, excitonic behaviour, and valley polarization properties of monolayer TMDs remains elusive to date.\\

The present work establishes high-energy (1 MeV) electron beam irradiation as a controlled method for defect engineering in monolayer MoS$_2$. We introduced defects through a tailored approach by varying the irradiation time, and systematically investigated their influence on electronic structure changes and optical properties, which have remained largely unexplored under such high energetic irradiation conditions. A detailed PL spectroscopy study is employed to understand the characteristics of defect states appearing at 200-300meV below the neutral exciton. Furthermore, we demonstrated the critical role of defects on the valley polarization properties of monolayer MoS$_2$ using circular polarization resolved PL measurement. Raman spectroscopy, X-ray photoelectron spectroscopy (XPS), and Kelvin probe force microscopy (KPFM) provide complementary insights into the structural, chemical, and electronic modifications post-irradiation. In addition, first-principles calculations were performed to corroborate the experimental observations. This work underscores a reliable approach for defect engineering in monolayer MoS$_2$ via high-energy electron beams and elucidates the interplay between irradiation parameters, defect formation, and excitonic behavior. These insights from the response of MoS$_2$ to harsh irradiation conditions pave the way for the development of advanced optoelectronic and quantum devices.

\section{Results and Discussions}
\begin{figure}[h!]
 \centering
  \includegraphics[width=13cm,height=11.31cm]{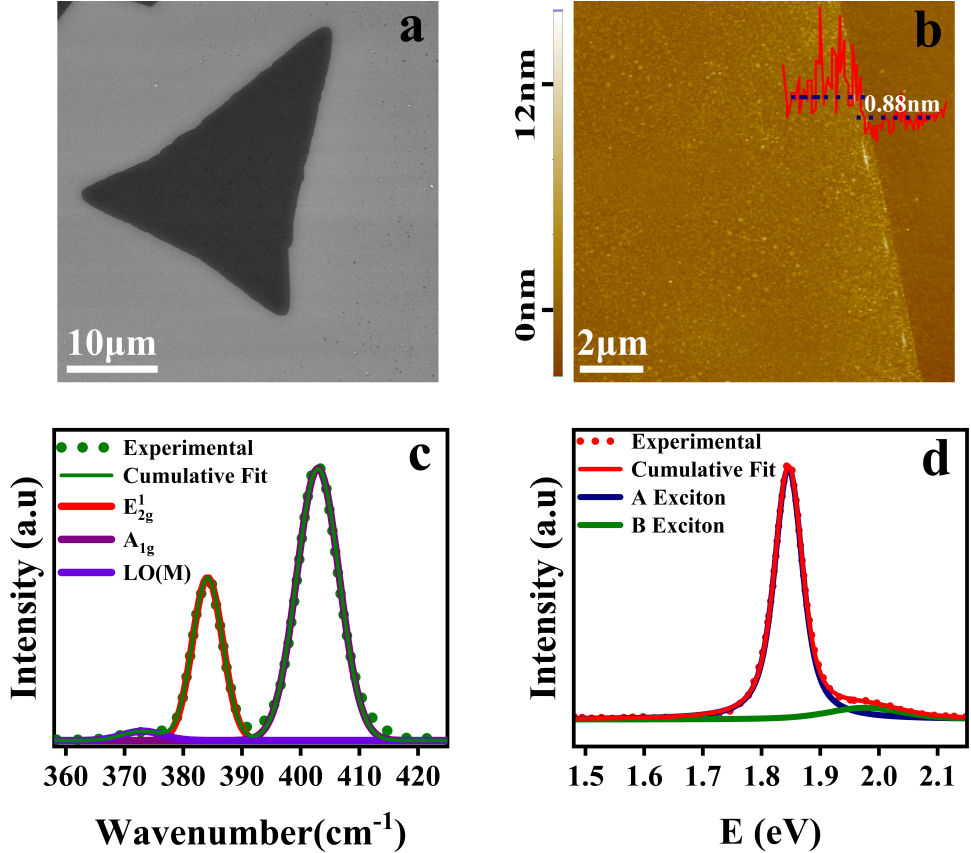}
  \caption{Characterization of CVD-grown MoS$_2$. (a) Scanning Electron Microscopy (SEM) image, (b) Atomic force Microscopy (AFM) image, (c) Raman spectrum, and (d) PL spectrum of monolayer MoS$_2$}
  \label{Figure1}
\end{figure}
Monolayer MoS$_2$ samples were prepared on SiO$_2$/Si substrate via the chemical vapor deposition (CVD) technique. The SEM image exhibits triangular domains of MoS$_2$ as shown in Figure~\ref{Figure1}(a). The SEM images taken from different regions of the sample, along with optical microscopy images, are included in Figure S.1 in the supporting information. The flake thickness of 0.88 nm, as obtained from the atomic force microscopy analysis (AFM) (Figure~\ref{Figure1}(b)), confirms the monolayer nature of the as-grown sample. Figures~\ref{Figure1}(c) and \ref{Figure1}(d) present the Raman and PL spectra of the MoS$_2$ flake, respectively. Gaussian fitting of the Raman spectrum displays two prominent modes at 384.13 cm$^{-1}$ $(E_{2g}^{1})$ and 403 cm$^{-1}$ (A$_{1g}$), separated by a frequency difference of 18.87cm$^{-1}$ which is consistent with the nature of monolayer MoS$_2$~\cite{66}. A subtle peak detected near 374cm$^{-1}$ is attributed to the longitudinal optical (LO(M)) mode arising from the sulfur vacancy defects in the MoS$_2$ lattice~\cite{67}. The PL spectrum shows a strong excitonic emission that can be resolved into an A exciton at $\sim$1.85 eV, indicating the direct recombination at the K point. A weak shoulder at $\sim$1.98 eV arises from the transition between the conduction band and the spin-split valence band. The obtained PL signatures further support the monolayer behaviour of MoS$_2$ flake~\cite{14}. \\
\begin{figure}[h!]
  \centering
  \includegraphics[width=0.98\linewidth]{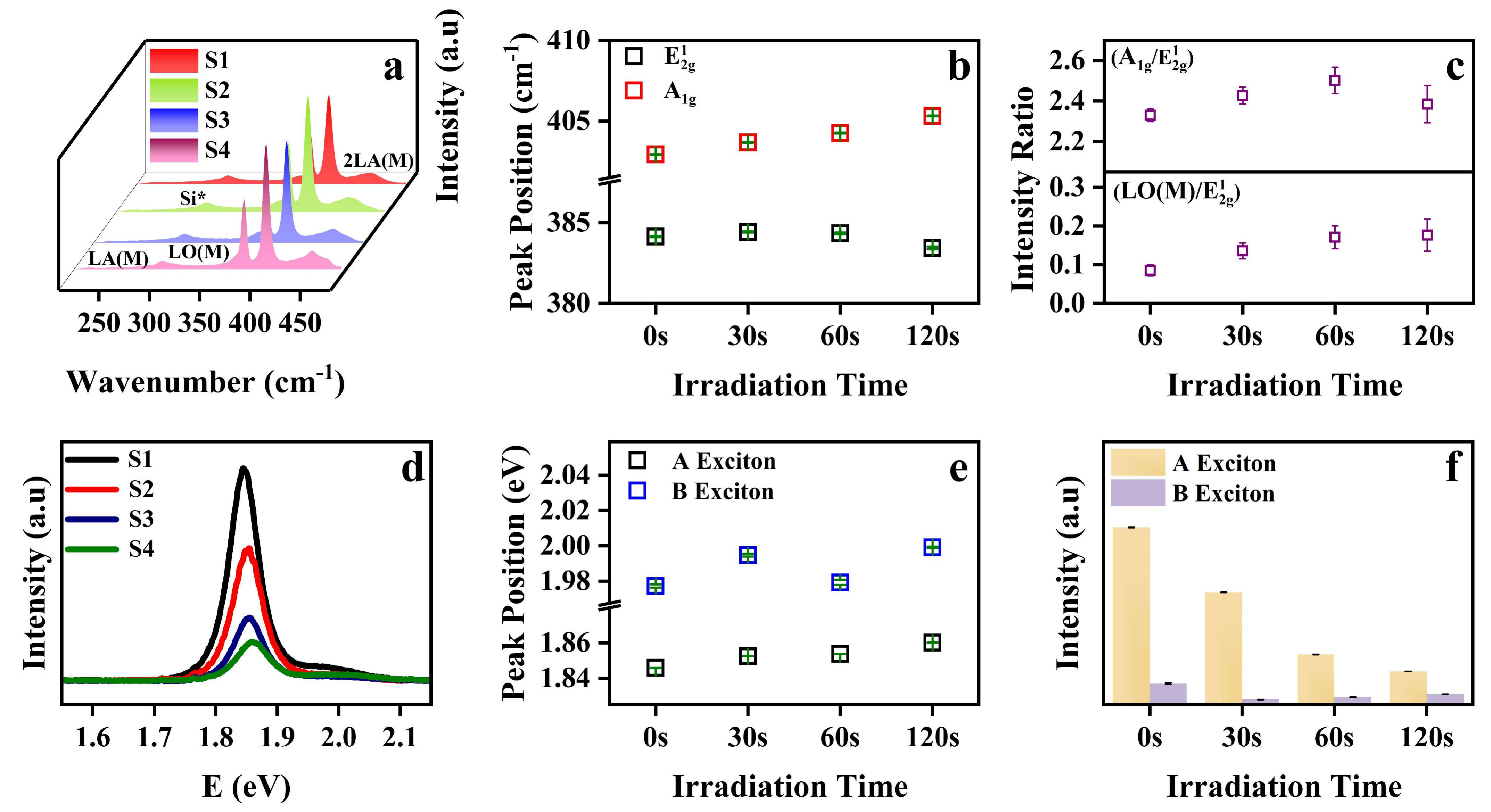}
  \caption{ High energy (1MeV) electron-beam irradiated MoS$_2$ at different irradiation times. (a) Raman spectra, (b) Extracted peak positions of E$_{2g}^{1}$ and A$_{1g}$ mode, (c) Intensity ratios: $\frac{A_{1g}}{E_{2g}^{1}}$ and $\frac{LO(M)}{E_{2g}^{1}}$, (d) PL spectra, (e) Extracted peak positions of A and B excitons and (f) Corresponding peak intensities.}
  \label{Figure2}
\end{figure}

The CVD-grown monolayer MoS$_2$ samples were exposed to a 1 MeV electron beam with a flux of $3\times10^{6} electrons/s$ for 30s, 60s, and 120s to examine the effect of high-energy electron beams on their structural and optical properties. The as-grown sample and those irradiated for 30s, 60s, and 120s are hereafter referred to as S1, S2, S3 and S4, respectively. No noticeable change in the optical contrast or surface morphology was observed in 
MoS$_2$ post-irradiation, as confirmed by the optical microscopy and SEM characterizations (Figure S.2 in supporting information). Figure~\ref{Figure2}(a) showcases the evolution of Raman spectra with respect to irradiation time. The $E_{2g}^{1}$ peak undergoes a slight blueshift till 60s of irradiation and a redshift by 1$cm^{-1}$ at 120s as illustrated in figure~\ref{Figure2}(b). In contrast, the A$_{1g}$ peak exhibits a gradual blue shift with increasing exposure time. Such Raman shifts are consistent with the formation of chalcogen vacancies. The observed redshift of the $E_{2g}^{1}$ mode indicates the dissociation of the Mo-S bond upon electron beam exposure, which reduces the number of Mo-S bonds participating in the in-plane vibration and weakens the force constant. On the other hand, the shift in A$_{1g}$ mode reflects an increased out-of-plane vibration and hence a larger force constant~\cite{68}. An enhancement in the intensity ratio between the A$_{1g}$ and E$_{2g}^{1}$ peak is also observed for irradiated samples(Figure~\ref{Figure2}(c)). Notably, the longitudinal optical mode LO(M) appearing at $\sim$374$cm^{-1}$ increases in intensity with electron exposure, denoting a rise in the defect density~\cite{69}. It also exhibits a progressive shift towards higher wavenumber with prolonged exposure. This behavior suggests phonon stiffening due to local bond strengthening near the vacancy sites. We also notice the occurrence of the longitudinal acoustic Raman mode, denoted by LA(M) in Figure~\ref{Figure2}(a), at $\sim$227cm$^{-1}$, which further signifies the presence of irradiation-induced disorder in the sample. The corresponding PL spectra are shown in Figure~\ref{Figure2}(d). A notable blueshift of the excitonic peaks with a significant quenching of PL intensity is evident in the PL spectrum following irradiation (Figure~\ref{Figure2}(e,f)). \\

\begin{figure}[h!]
  \centering
  \includegraphics[width=0.98\linewidth]{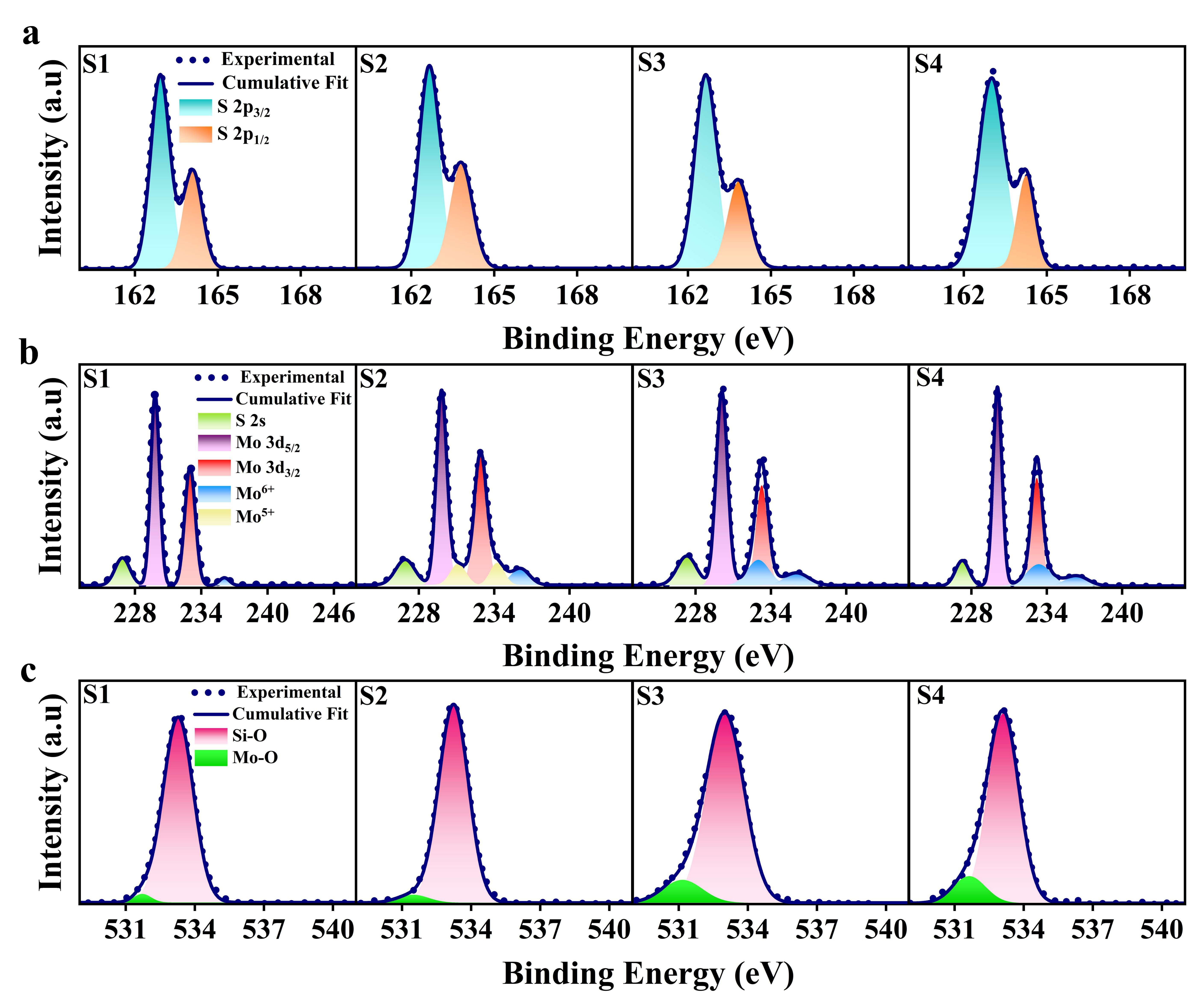}
  \caption{XPS characterization of as grown and irradiated MoS$_2$. The XPS spectra of: (a) S 2p core electrons, (b) Mo 3d core electrons, and (c) O 1s core electrons. The spectral changes validate the oxygen interaction at chalcogen vacancies.}
  \label{Figure3}
\end{figure}

XPS and KPFM measurements were conducted in order to attain a deeper understanding of the chemical state and electronic structure changes underlying these spectral variations. The Sulfur (S) 2p core level spectrum of sample S1 shown in Figure~\ref{Figure3}(a), exhibits the characteristic spin-orbit doublet at binding energies of $\sim$162.65 eV and $\sim$163.81 eV for the S 2p$_{3/2}$ and S 2p$_{1/2}$, respectively, consistent with the S$^{2-}$ state in MoS$_2$. Upon electron exposure for 30s, the spectral features of S 2p remain nearly intact and thus imply the minimal loss of sulfur at shorter irradiation times. However, a slight shift toward higher binding energy is visible with longer electron exposure. Figure~\ref{Figure3}b represents the XPS spectrum of the Molybdenum (Mo) 3d core electrons. The extracted peak positions at $\sim$229.83eV and $\sim$232.98eV of sample S1 correlates to the Mo$^{4+}$ 3d$_{5/2}$ and 3d$_{3/2}$ binding energy states, respectively. The distinct peak at 226.87 eV is assigned to the S 2s level and the weaker XPS signal at 236.10 eV denotes the Mo$^{6+}$ oxidation state~\cite{70}. It is worth noting that the Mo$^{4+}$ doublet also exhibits a similar trend in binding energy shifts as seen for S 2p. The S 2p and Mo 3d binding energies are summarized in Table S.1 of the supporting information. XPS analysis of sample S2 revealed that 30s of electron beam exposure leads to the appearance of a new doublet at 231.13 eV (3d$_{5/2}$) and 234.26 eV (3d$_{3/2}$) in the Mo 3d spectra, signifying the emergence of the Mo$^{5+}$ state after irradiation~\cite{71}. We point out that the Mo$^{6+}$ becomes more pronounced with continuous exposure to the electron beam with its doublet peaking at $\sim$233eV (3d$_{5/2}$) and $\sim$236eV (3d$_{3/2}$) in samples S3 and S4.  The increase in intensity of the Mo$^{6+}$ state with electron exposure points to the severity of oxidation due to the formation of a larger number of sulfur vacancy sites. These effects stem from the higher incident energy of the electron beam exceeding the threshold displacement energy of the Mo and S atoms. Electronic excitations can be dominant at lower irradiation times and thereby weaken the Mo-S bond. This facilitates the incorporation of oxygen into the electronically activated Mo atoms which culminates in the formation of the Mo$^{5+}$ state in sample S2 (Figure~\ref{Figure3}(b)). Meanwhile, prolonged electron beam exposure induces more sulfur vacancy defects by knock-on damage. This leaves the Mo centres undercoordinated. Consequently, the active sites of the sulfur vacancies react with oxygen in the ambient and result in the formation of a Mo$^{6+}$ state in the Mo 3d spectra. It is obvious from the Mo 3d and S 2p spectra that the selective oxidation of Mo occurs, and the S remains largely unaffected. Given the higher electronegativity of the oxygen atom, this oxidation process decreases the electron density around Molybdenum, which in turn leads to the observed upward shift in the binding energies. These observations are further corroborated by the O 1s spectra depicted in Figure~\ref{Figure3}c. The broad intense peak at 533.3 eV is attributed to the Si-O bonding from the substrate, and the peak at 531.7 eV corresponds to the Mo-O bond. Additionally, we observed an increase in the intensity of the Mo–O component in the O 1s spectrum following electron beam irradiation. This observation is consistent with the irradiation-induced increase of the Mo$^{6+}$ signal in the Mo 3d spectra. Thus, the XPS findings emphasize that the observed structural changes emanate from the irradiation-induced defects followed by subsequent oxidation.\\
\begin{figure}
 \centering
  \includegraphics[width=0.98\linewidth]{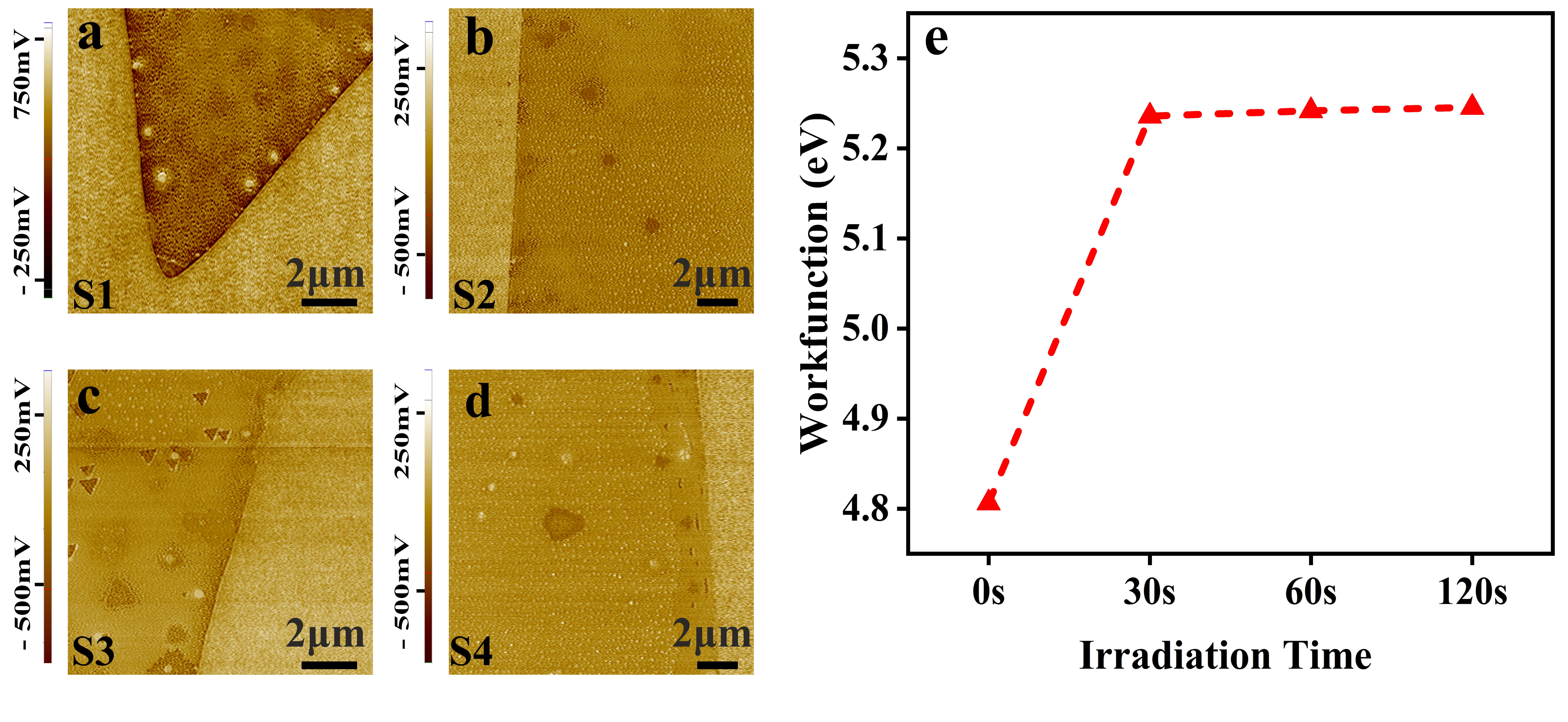}
  \caption{KPFM Analysis. Surface potential mapping of (a) as-grown MoS$_2$ and (b)-(d) irradiated samples. (e) Calculated work function as a function of irradiation time. With prolonged electron exposure, surface potential decreases, and the corresponding increase in work function signifies p-type doping.}
  \label{Figure4}
\end{figure}

We further analyzed the surface electronic changes that occurred to monolayer MoS$_2$ upon electron beam irradiation using the KPFM technique. This technique enables us to probe the surface potential variations directly correlating to the work function of the sample. The surface potential mapping of the as-grown and irradiated flakes is shown in Figures~\ref{Figure4}(a-d). The work function of the tip was calibrated using HOPG as the reference prior to the measurement. The work function of the sample can be calculated using the formula~\cite{72}
\begin{equation}
   \phi_{\text{sample}} = \phi_{\text{tip}} -  eV_{\text{CPD}}
\end{equation}
where $\phi_{\text{sample}}$ and $\phi_{\text{tip}}$ indicate the work function of the sample and tip, respectively, and V$_{CPD}$ is the contact potential difference between the tip and the sample. From Figures~\ref{Figure4}(a)-(d), we have observed that the surface potential of MoS$_2$ samples decreases with irradiation exposure. Thus, a notable rise of 0.4eV is observed in the work function of the sample after irradiation, with an increase from 4.8eV for the as-grown sample to 5.24eV in the 120s irradiated sample (Figure~\ref{Figure4}(e)). The corresponding work function mapping is shown in Figure S.3 of the supporting information. The increase in work function points to the p-type doping behaviour induced by e-beam irradiation. The sulfur vacancy sites formed in MoS$_2$ as a result of electron beam irradiation facilitate the oxygen incorporation from the ambient atmosphere, as evidenced by the XPS results. Owing to the higher electronegativity of the oxygen atom, the oxygen withdraws electrons from the MoS$_2$ lattice~\cite{79}. The oxygen incorporation and the subsequent charge transfer cause the Fermi level to shift to the valence band edge, thereby resulting in an increase in work function. Samples irradiated for longer periods tend to have increased vacancies and more oxidation, and, therefore, the highest work function. This further correlates with the increased Mo$^{6+}$ oxidation signature observed in the XPS results post-irradiation.\\

\begin{figure}[h!]
  \centering
  \includegraphics[width=0.98\linewidth]{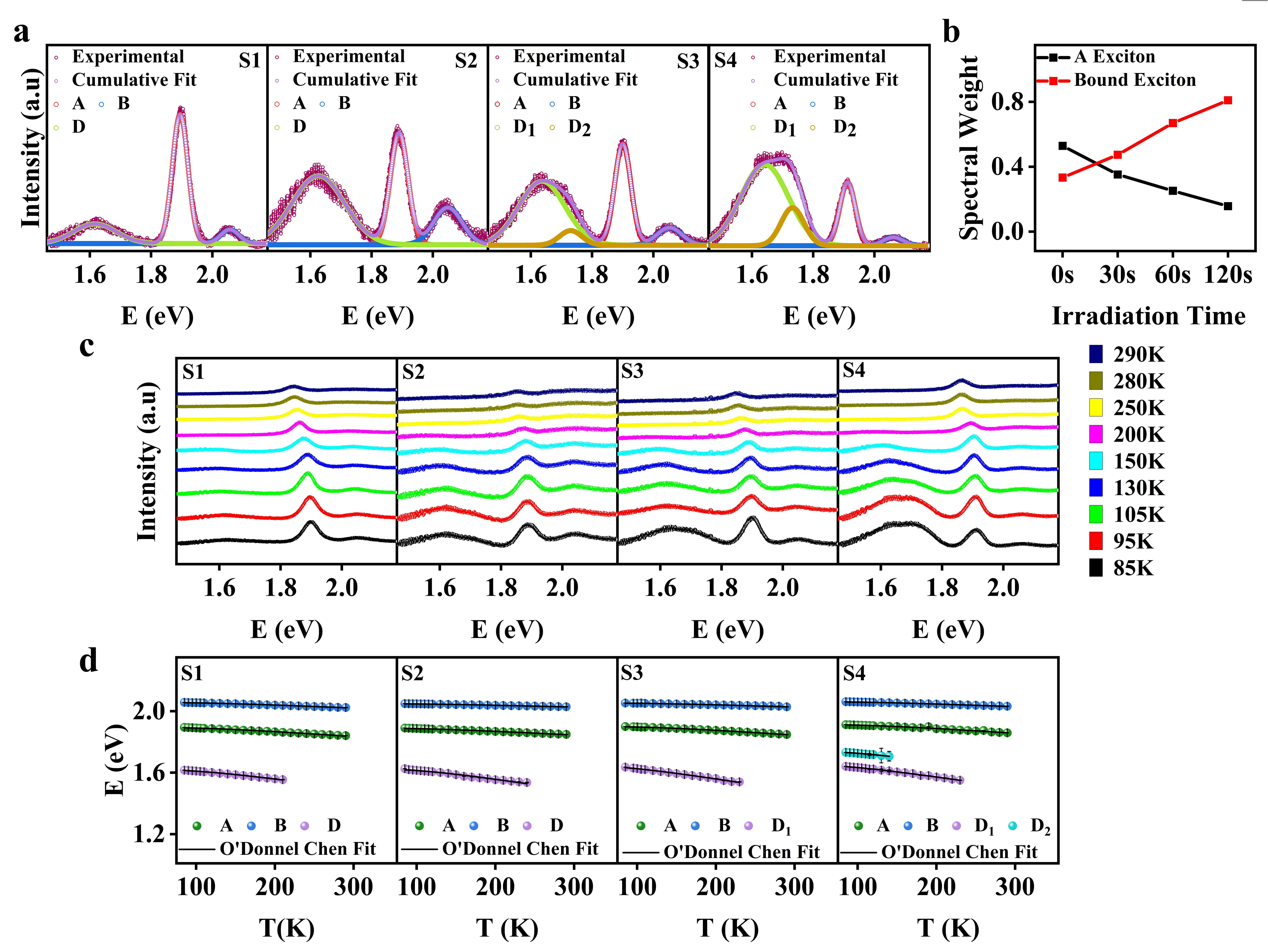}
  \caption{Temperature-dependent PL Spectra. (a) PL spectra of as-grown and irradiated samples recorded at 85K, (b) Spectral weight of the samples as a function of irradiation time, (c) PL spectra procured at different temperatures. (d) Extracted peak positions as a function of temperature. The solid black line indicates the fitting curve using the O'Donnell-Chen equation.}  
  \label{Figure5}
\end{figure}

Temperature-dependent PL studies were employed to correlate the observed chemical state modifications and doping characteristics with the excitonic properties. Figure \ref{Figure5}(a) show the PL spectra of as-grown and irradiated samples acquired at 85 K. The recorded PL spectra were well fitted using the Gaussian function. Apart from distinct A and B excitonic transitions, a broad defect-related peak, labelled as D, appears at 200-300 meV below the A exciton transition. The weak D emission at $\sim$1.62 eV is apparent in sample S1, which denotes the defect-induced deep mid-gap states that are likely caused by the native or oxygen-bound sulfur vacancies as indicated from the Mo-O bonding signatures in
XPS spectra. This defect-related peak (D) broadens and becomes more intense with prolonged exposure to electrons, denoting an increased density of defects in MoS$_2$ lattice after irradiation. Further, deconvolution of the D peak in S3 and S4 unveils two spectral peaks, D$_1$ and D$_2$, centred at  $\sim$1.64 eV and $\sim$1.73 eV, respectively. These states also arise from the oxygen interaction with S vacancies with deeper defect states at $\sim$1.64eV and shallower defect states at $\sim$1.73eV. The simultaneous appearance of both peaks signals a rise in defect levels within the bandgap and thus an increase in defect concentration in MoS$_2$ with increasing irradiation time. It is noteworthy that the spectral weight of defect-bound emission increases with irradiation time, whereas that of A exciton decreases (Figure~\ref{Figure5}b). This increase in defect density weakens the free exciton recombination by activating the non-radiative recombination channels and through defect-mediated recombination. Consequently, there is a reduction in the spectral weight of the A exciton at 85K and a decrease in PL intensity at room temperature following irradiation(Figure~\ref {Figure2}d). These observations on the enhancement of the defect-related emission with irradiation time are in line with the observed rise of the Mo$^{6+}$ peak in Mo 3d XPS spectra and the Mo-O bonding peak found in the O 1s XPS spectra after irradiation. Figures~\ref{Figure5}(c) represent the variation in PL intensity with temperature using a laser power of 0.230 mW. Owing to the non-radiative recombination channels, the bound exciton emission weakens with temperature and persists up to $\sim$210K-240K, emphasizing strong carrier localization at the trap level. The spectrum is solely dominated by A exciton emission at higher temperatures. In addition, the disappearance of D$_2$ peak above 140K further elucidates its shallow behavior. Hence, the weakly bound nature of D$_2$ exciton makes it more susceptible to non-radiative recombination channels. It is worth noting that the weak D$_2$ emission is detected only at 85K in sample S3 (Figure \ref{Figure5}c), probably due to thermal quenching or de-trapping resulting from carrier transfer to deeper defect states. Figure \ref{Figure5}(d) portrays the corresponding variation in excitonic peak positions with temperature. The redshift observed in the peak positions is analyzed using the O'Donnell and Chen equation as given below~\cite{73}:
\begin{equation}
E(T) = E(0) - S \langle \hbar\omega \rangle \left[ \coth\left( \frac{ \langle \hbar\omega \rangle }{2k_B T} \right) - 1 \right]
\end{equation}
where  E(0) denotes the exciton peak energy at 0K, $\langle \hbar\omega \rangle$ is the average phonon energy, and S is the Huang-Rhys factor referring to the exciton-phonon interaction. Fitting parameters are shown in Table.\ref {Table1}. The peak energy of the A exciton at 0K, denoted by E(0), shows a progressive blueshift from 1.894 eV to 1.913 eV with increasing irradiation time. A similar trend is also noticed in the room temperature PL spectra, confirming the robustness of excitonic behaviour across various temperature regimes. The extracted average phonon energy $\langle \hbar\omega \rangle$ falls in the range of 31 - 33 meV for the A exciton which agrees well with the characteristic phonon energies of E$_{2g}^{1}$ and A$_{1g}$ modes of MoS$_2$ and the accompanying Huang-Rhys factor ranges from $\approx$ 1.6 to 2.2\cite{78}, indicating the weaker coupling of A excitons with phonons. The B exciton peak also displays comparable average phonon energy values and a lower Huang-Rhys factor (0.8-1.6), which further substantiates the weak electron-phonon interaction typical of direct excitons. Meanwhile, the E(0) value of the deep defect-bound exciton exhibits a peak shift from $\approx$1.62 eV to 1.65 eV. The Huang Rhys factor for this exciton spans from $\approx$4.5 to 5 and the notably high value confirms stronger electron-phonon coupling on account of the localized nature of defect-bound excitons. Given that the average phonon energies are $\approx$ 22 - 28 meV, it can be inferred that the D exciton couples to phonons of low-frequency vibrational modes. Similarly, shallow defect state (D$_2$) observed in sample S4 has an average phonon energy of 27 meV and an S factor of 4.8, further highlighting the strong coupling between defect-bound excitons and phonons.\\

\begin{table}[htbp]
  \caption{Parameters ($E_g(0)$, $S$, $\langle \hbar\omega \rangle$) extracted by fitting the temperature-dependent exciton peak positions using O'Donnel and Chen's equation.}
  \label{Table1}
  \centering
  \resizebox{\textwidth}{!}{%
    \begin{tabular}{l c c c c c c c c c c c c}
      \hline
      \textbf{Irradiation Time} 
      & \multicolumn{3}{c}{\textbf{A exciton}} 
      & \multicolumn{3}{c}{\textbf{B exciton}} 
      & \multicolumn{3}{c}{\textbf{D1 exciton}} 
      & \multicolumn{3}{c}{\textbf{D2 exciton}} \\
      
      & $E_g(0)$ (eV) & $\langle \hbar\omega \rangle$ (eV) & $S$
      & $E_g(0)$ (eV) & $\langle \hbar\omega \rangle$ (eV) & $S$
      & $E_g(0)$ (eV) & $\langle \hbar\omega \rangle$ (eV) & $S$
      & $E_g(0)$ (eV) & $\langle \hbar\omega \rangle$ (eV) & $S$ \\
      \hline

      0 s & 
      $1.894 \pm 0.001$ & $0.032 \pm 0.003$ & $2.25 \pm 0.16$ &
      $2.057 \pm 0.001$ & $0.031 \pm 0.003$ & $1.37 \pm 0.09$ &
      $1.620 \pm 0.002$ & $0.028 \pm 0.003$ & $4.46 \pm 0.39$ &
      -- & -- & -- \\

      30 s & 
      $1.889 \pm 0.001$ & $0.031 \pm 0.003$ & $1.67 \pm 0.10$ &
      $2.049 \pm 0.000$ & $0.031 \pm 0.003$ & $0.81 \pm 0.05$ &
      $1.625 \pm 0.004$ & $0.028 \pm 0.005$ & $5.02 \pm 0.59$ &
      -- & -- & -- \\

      60 s & 
      $1.901 \pm 0.000$ & $0.033 \pm 0.001$ & $2.20 \pm 0.06$ &
      $2.053 \pm 0.001$ & $0.032 \pm 0.007$ & $1.00 \pm 0.14$ &
      $1.645 \pm 0.004$ & $0.022 \pm 0.003$ & $4.96 \pm 0.34$ &
      -- & -- & -- \\

      120 s & 
      $1.913 \pm 0.002$ & $0.032 \pm 0.006$ & $2.17 \pm 0.26$ &
      $2.061 \pm 0.001$ & $0.031 \pm 0.003$ & $1.19 \pm 0.07$ &
      $1.650 \pm 0.003$ & $0.022 \pm 0.002$ & $4.66 \pm 0.25$ &
      $1.738 \pm 0.002$ & $0.027 \pm 0.004$ & $4.80 \pm 0.64$ \\
      \hline
    \end{tabular}
  }
\end{table}

\begin{figure}[h!]
  \centering
  \includegraphics[width=0.98\linewidth]{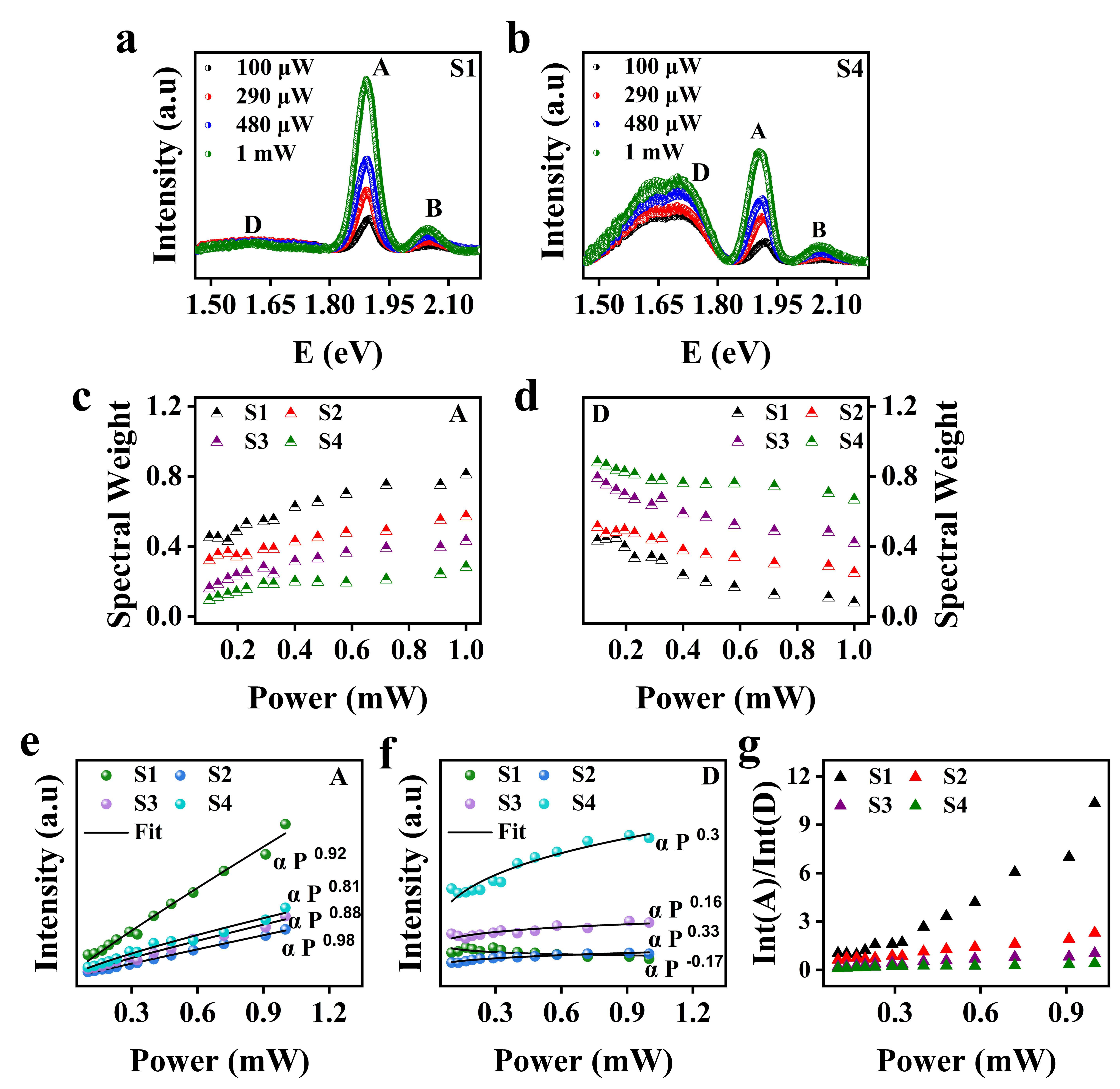}
  \caption{Excitation power-dependent PL spectra. PL spectra of (a) as-grown (S1) and (b) 120s irradiated sample (S4) at various excitation powers. Spectral weight as a function of laser power for the (c) A exciton and (d) D exciton, PL intensity of (e) A exciton and (f) D exciton as a function of excitation laser power. The black solid curve indicates the fitting using the power law: I $\alpha$ P$^k$. (g) Intensity ratio of the A exciton to the D exciton as a function of laser power.}
  \label{Figure6}
\end{figure}

To further validate the bound nature of defect states induced by the oxygen incorporation at sulfur vacancy sites, PL measurements were performed at different excitation laser powers ranging from 0.1mW - 1.0mW. PL spectra at excitation powers of 0.1 mW, 0.29 mW, 0.48 mW and 1 mW for samples S1 and S4 are shown in Figures \ref{Figure6}(a) and \ref{Figure6}(b), respectively. The spectra acquired at intermediate excitation powers and the power-dependent PL spectra of samples S2 and S3 are included in Figure S.5 of the supporting information. It can be noticed from Figures \ref{Figure6}(c) and \ref{Figure6}(d) that the spectral weight of A exciton increases with an increase in laser power, whereas that of D exciton decreases. To understand the interplay between the free and defect-bound exciton, we analyzed the integrated PL intensity as a function of excitation laser power using the power law of the form, I $\alpha$ P$^k$~\cite{74}. 
Figures \ref{Figure6}(e) and \ref{Figure6}(f) display the dependence of the intensity of A and D excitons on laser power. The obtained value of k $\approx$ 0.92 for the A exciton in sample S1 suggests a nearly linear dependence of intensity on excitation power (Figure \ref{Figure6}e), implying that the delocalized excitons dominate the recombination process. Meanwhile, the corresponding defect-bound exciton exhibits a negative power-law relation (Figure \ref{Figure6}(f)) with k $\approx$ -0.17, reflecting the quenching of defect emission with the increase of laser power. This negative value of the power law constant is likely due to the lower density of available trap states, which become fully occupied even at relatively low excitation power. The decrease in the power law exponent of A exciton to 0.8 in sample S4 (Figure \ref{Figure6}(e)) signifies enhanced non-radiative recombination due to irradiation-induced defects in the MoS$_2$ lattice. In contrast, a sublinear relationship is observed for the bound exciton post-irradiation with k values ranging from 0.1 - 0.3 as shown in Figure \ref{Figure6}(f), which implies the saturation tendency of these defect states at higher laser power. This further confirms the emergence of a higher density of optically active defect states following irradiation. The results are further supported by the evolution of the A/D intensity ratio with excitation power as depicted in Figure \ref{Figure6}(g). The as-grown sample exhibits a superlinear increase with laser power, which is in line with the dominance of free exciton and limited availability of defect states as aforementioned. A notable reduction in the A/D ratio reaffirms the growing influence of defect-mediated recombination upon irradiation. The defect emission remains significant despite its tendency to saturate at higher laser powers owing to the increased density of defects within the sample. This is further evidenced by the D exciton emission prevailing over the A exciton in samples S3 and S4, even at high laser power(Figure \ref{Figure6}(g)).\\

\begin{figure}[h!]
  \centering
  \includegraphics[width=0.98\linewidth]{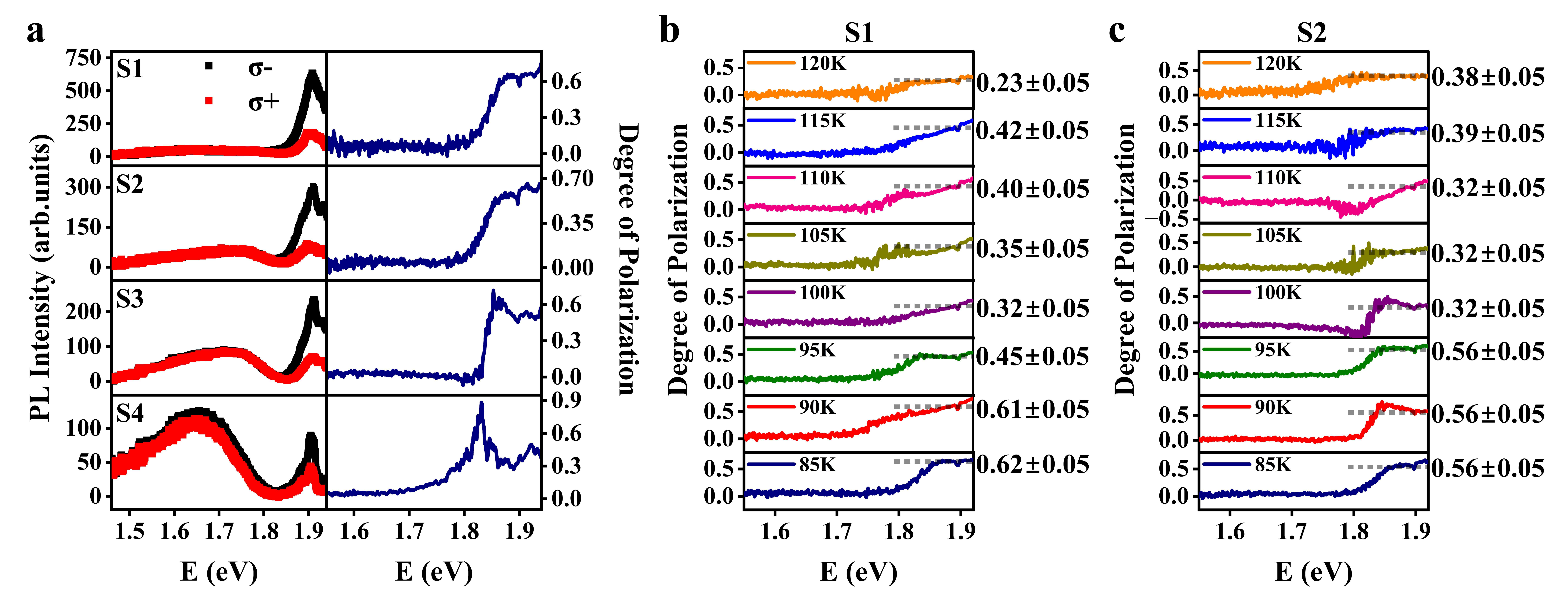}
  \caption{Polarization resolved PL Spectra. (a) Circularly polarized PL spectra corresponding to $\sigma$- excitation at 85K. The corresponding temperature dependence of the degree of circular polarization of (b) as-grown, and (c) 30s irradiated sample.}
  \label{Figure7}
\end{figure}

The impact of electron beam irradiation on excitonic behavior was further examined using circular polarization-resolved PL measurements. Circularly polarized excitation in monolayer MoS$_2$ selectively drives excitons into one of the two inequivalent K valleys, giving rise to valley polarization~\cite{12,13}. Figure \ref{Figure7}(a) exhibits the polarization-resolved PL spectra obtained at 85K for as-grown and irradiated samples under $\sigma-$ excitation (left circularly polarized light). The corresponding degree of polarization (DOP) given by $DOP = \frac{I^- - I^+}{I^- + I^+}$, wherein $I^-$ and $I^+$ indicate the measured PL intensity corresponding to $\sigma-$ (left circularly polarized) and $\sigma+$ (right circularly polarized) light, respectively, is displayed alongside the spectra. It is evident from
Figure 7a that the polarization is present solely in the A excitonic feature at $\sim$1.9eV while the D peak remains unpolarized. Furthermore, a substantial reduction in valley polarization is seen with the increase in defect peak intensity. The calculated degree of polarization drops from 62\% in sample S1 to 38\% in sample S4. This decrease in valley polarization can be attributed to the pronounced defect-mediated intervalley transition process and carrier trapping at the mid-gap defect levels. Figures \ref{Figure7}(b) and \ref{Figure7}(c) indicate the temperature dependence of valley polarization of samples S1 and S2, respectively. It is important to note that the valley polarization decreases with increasing temperature. The observed non-monotonic variation in valley polarization can be understood within the framework of Maialle Silva Sham (MSS) mechanism~\cite{75,76}, which takes into account the long-range electron-hole exchange interaction. Figure \ref{Figure7}(b) depicts that the valley polarization of the A exciton reduces from $\sim$62\% at 85K to $\sim$32\%  at 100K in the as-grown sample. Owing to the relatively lower density of defects, the sample is initially in the weaker scattering regime. Here, the observed decline in valley polarization is facilitated by the increased thermal activation of low-energy acoustic phonons, which raises the momentum scattering rate of excitons. Since the material is driven into a strong scattering regime upon raising the temperature further, the exchange-induced precession weakens and causes a slight gain in polarization to 40\%. Beyond 120K, valley polarization is further quenched, likely due to the enhanced phonon-mediated (optical and acoustic) intervalley scattering. The sample S2 (Figure \ref{Figure7}(c)) exhibits a lower initial polarization (~56\% at 85 K) as a result of the enhanced effect of defect levels upon irradiation. It also possesses a comparable non-monotonic variation with temperature consistent with the MSS mechanism. Furthermore, the phonon-driven features that are easily discernible in the as-grown sample are obscured by the persistent defect-induced depolarization channels. In summary, our findings emphasize that the increase in defect density suppresses valley polarization, and the circular polarization-resolved PL measurements are an effective probe to quantify this effect. \\

\begin{figure}[h!]
  \centering
  \includegraphics[width=0.98\linewidth]{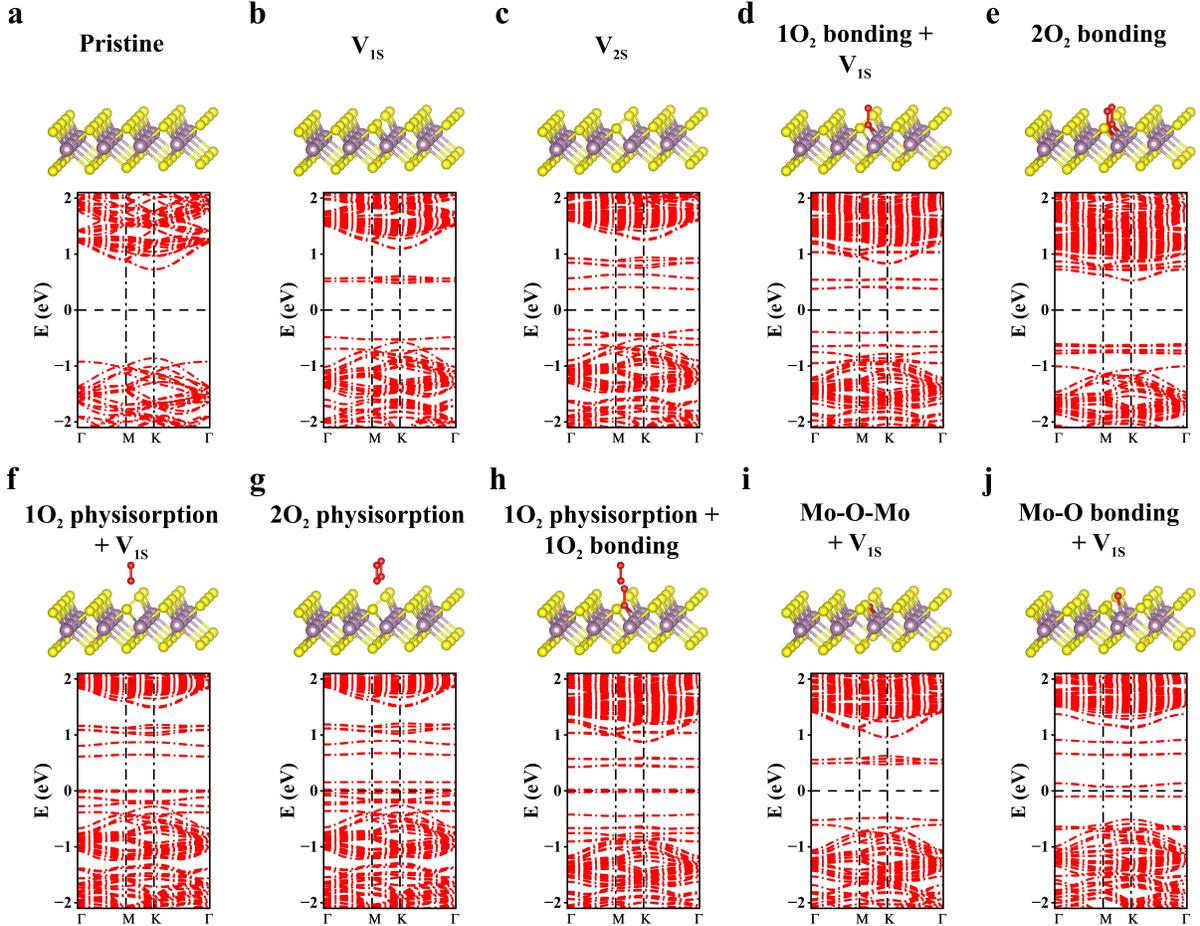}
  \caption{The band structure of 4x4x1 supercell of MoS$_2$ for different configurations.(a) pristine MoS$_2$,  (b) with a single sulfur vacancy, (c) with disulfur vacancies, (d) O$_2$ bonding at the sulfur vacancy site along with an a sulfur vacancy, (e) two O$_2$ bonding at the sulfur vacancy sites, (f) O$_2$ physisorption at the sulfur vacancy site along with a sulfur vacancy, (g) two O$_2$ physisorption at the sulfur vacancy sites (h) O$_2$ physisoprtion and bonding at the sulfur vacancy site (i,j) Mo-O-Mo bridge and Mo-O bonding with a sulfur vacancy, respectively.} 
  \label{Figure8}
\end{figure}

Through first-principle calculations, we examine how various oxygen interactions at sulfur vacancies modify the electronic band structure of MoS$_2$ in order to understand the emergence of the defect peak in the PL spectra. The band structure calculations are performed on a 4x4x1 supercell of monolayer MoS$_2$. Figure \ref{Figure8}(a) represents the
band structure of the pristine MoS$_2$ monolayer. The direct band gap nature is evidenced at the K point. Addition of monosulfur and disulfur vacancies creates defect levels closer to the conduction band (Figures \ref{Figure8}(b) and \ref{Figure8}(c)). More donor states appear near the conduction band with increasing sulfur vacancy concentration, resulting in n-type doping of MoS$_2$(See Figure S.6 in supporting information). We further consider the interaction of oxygen with these sulfur vacancy sites to comprehend the Mo-O peak appearing in the XPS spectra. Figure \ref{Figure8}(d) presents the band structure corresponding to O$_2$ bonding at the sulfur vacancy site in the presence of a nearby sulfur vacancy. It is clear from Figure \ref{Figure8}(d) that the defect states arising from the O$_2$ bonding appear near the valence band, indicating their acceptor-like nature. An increase in the number of O$_2$ bonds results in more acceptor-like states closer to the valence band (see Figure \ref{Figure8}(e)). We have also explored the possibility of oxygen physisorption at the Sulfur vacancy sites. Figures \ref{Figure8}(f) and \ref{Figure8}(g) illustrate the O$_2$ adsorption at the sulfur vacancy site accompanied by a sulfur vacancy and two O$_2$ adsorption at the sulfur vacancy sites, respectively. These configurations also introduce defect states near the valence band edge. Our calculations further predict that the combined presence of O$_2$ physisorption and O$_2$ bonding at sulfur vacancy sites also leads to the formation of acceptor states(Figure \ref{Figure8}(h)). Even though the donor states from the sulfur vacancies appear in the above scenarios, the acceptor states lie closer to the band edge, causing easier ionization. Thus, the shallow and deep acceptor levels formed near the valence band induce net p-type doping. These theoretical predictions are consistent with the obtained KPFM results. Additionally, Mo-O bonding also supports p-type doping, whereas the Mo-O-Mo bridge produces donor-like states, as shown in Figure \ref{Figure8}(i) and Figure \ref{Figure8}(j), respectively. The transition between the conduction band and these acceptor levels leads to the defect peak obtained in the PL spectra (Figure \ref{Figure5}(a)). It is also clear that energy separation between $\Gamma$ and K valley decreases with the addition of vacancies and oxygen interactions (refer Table S.2 in supporting information). This enhances the valley depolarization between the K and K' via $\Gamma$ through the phonon-mediated scattering process. It could further justify the experimentally observed reduction in valley polarization following electron exposure. 

\section{Conclusion}
In summary, we report a novel and controllable approach to engineer defects in monolayer MoS$_2$ through high-energy electron beam irradiation. The defect density was precisely tuned by varying the electron exposure time and keeping the electron beam energy and flux constant. The impact of high-energy electron exposure on monolayer MoS$_2$ was systematically probed using XPS, KPFM, Raman, and PL spectroscopy, which was further supported by first-principles calculations. Defect-bound emission intensity in the PL spectra increased with irradiation time, indicating a higher defect density in irradiated samples. This is further accompanied by the quenching of valley polarization. The emergence of Mo-O bonding and the p-type behavior in monolayer MoS$_2$ after irradiation was revealed from XPS and KPFM measurements. First-principles calculations mimicking the effects of these chemical and electronic modifications confirm that the observed defect peak arose from the transition between the conduction band and acceptor levels. Our findings demonstrate the effectiveness of the high-energy electron beam in the controllable introduction of defect states in monolayer MoS$_2$, and the methodology can be extended to other TMDs with proper optimization of irradiation parameters.

\section{Methods}
\subsection{CVD growth of \texorpdfstring{MoS$_2$}{MoS2}} MoS$_2$  was grown on a SiO$_2$/Si substrate via CVD technique using a horizontal tube furnace with two heating zones~\cite{77}. The growth substrate underwent sonication in acetone, isopropyl alcohol (IPA), and distilled water (DI) for 4 minutes each, and then it was blow-dried using Nitrogen. The cleaned substrate was placed in a face-down position on top of an Alumina boat containing 6mg of MoO$_3$ (Mo precursor) and kept in the second heating zone. The first heating zone contained 200mg of sulfur powder (Sulfur source) in another boat positioned $\sim$60cm upstream from the Mo precursor. We began by purging the tube with 300sccm of high-purity Argon gas to create an inert environment inside the quartz tube. The Mo and S zones were heated to 750$^{\circ}$C and 180$^{\circ}$C, respectively, in 50 minutes and held at those temperatures for 10 minutes. Argon was used as the carrier gas with a flow rate of 120sccm till 700$^{\circ}$C of the second heating zone and then dropped to 60sccm. The system was left to naturally cool down to room temperature following the growth process.\\

\subsection{Electron beam irradiation}Electron beam irradiation experiments were performed using the beamline of the compact THz Free Electron Laser facility at IUAC, New Delhi. The electron beam was produced in a 2.6-cell RF photocathode gun operating at 2860MHz via photoemission from a copper cathode illuminated by UV laser pulses. For the experiment, the irradiation energy was set to 1 MeV and the flux to $\sim 3 \times 10^{6}$ electrons/second. The electron incidence was normal to the MoS$_2$ surface.\\ 

\subsection{Photoluminescence Spectroscopy} PL measurements were performed on the samples at cryogenic temperature using a liquid-nitrogen cryostat. The measurements were conducted in backscattering geometry utilizing a microscope setup with a 50x long working distance objective. PL measurements were performed by means of a 532nm laser excitation with a power of 230$\mu$W. Circularly polarized excitation ($\sigma+/\sigma-$) was obtained by passing the linearly polarized beam from a He-Ne laser of 632nm wavelength through a quarter-wave plate. To resolve the $\sigma+$ and $\sigma-$ components of the emitted photons, another achromatic quarter-wave plate together with a Glan Taylor analyzer was mounted before the spectrometer's entrance slit. Spectral data were acquired with a monochromator (focal length~0.55m) coupled to a Peltier cooled CCD detector~\cite{76}.\\

\subsection{Raman Spectroscopy, FESEM, XPS, AFM and KPFM characterization} Raman spectroscopy was performed using a micro-Raman facility from Renishaw Invia Reflex (UK) with a 532nm excitation laser at a power of 50mW and a 2400lines/mm diffraction grating. FESEM images were taken with a TESCAN MIRA II at an accelerating voltage of 15 keV and detected using a secondary electron detector. The XPS characterization was done using AXIS Supra (Kratos analytical) with a monochromatic Al-K$\alpha$ X-Ray source of energy 1486.6eV. Park NX 10 model was used to carry out the AFM and KPFM measurements to obtain the surface topography and workfunction. The measurements were carried out in non-contact mode using a Cr/Au-coated Si tip. The tip workfunction was calibrated with respect to HOPG sample prior to the measurements. The AC voltage was applied to the tip with an amplitude of 3V and 4kHz, and the scan rate was maintained at 1Hz.\\

\subsection{Computational Details} The first-principle calculations were performed using Quantum Espresso package. A 4x4x1 supercell of monolayer MoS$_2$, with 10 vacuum, was chosen for all the calculations. Electron-ion interaction was modeled using projector-augmented-wave pseudopotential, whereas the exchange-correlation energy was accounted for using the Perdew-Burke-Ernzerhof (PBE) functional within the generalized gradient approximation (GGA). The effect of spin-orbit coupling was included in the calculations. Considering the kinetic cut off as 60 Ry, the system was fully relaxed using Broyden Fletcher–Goldfarb–Shanno (BFGS) algorithm until the force and energy were below $7.7 \times 10^{-2}$ Ry/Bohr and $1 \times 10^{-3}$ Ry, respectively. A 4x4x1 Monkhorst-Pack k-point grid was used to sample the Brillouin zone for geometry relaxation and band structure calculations. 
\section*{Acknowledgments}
A. G. and J. M. acknowledge the Indian Institute of Technology Hyderabad for providing the research facilities. A. G. and J. M. also acknowledge the FEL group, IUAC, New Delhi, for providing the facility to carry out electron beam irradiation experiments. A. G. acknowledges the Ministry of Education (MoE), India, for funding support. The authors acknowledge the National Supercomputing Mission (NSM) for providing computing resources of ‘PARAMSEVA’ at IIT, Hyderabad, which is implemented by C-DAC and supported by the Ministry of Information Technology (MeitY) and Department of Science and Technology (DST), Government of India.

\section*{Data Availability}
The data that support the findings of this study are available from the corresponding author upon reasonable request.

\section*{Conflict of Interest}
The authors declare no conflict of interest.

\end{document}